\documentclass[useAMS,usenatbib,usegraphicx]{mn2e}
\usepackage{mathrsfs}
\usepackage{amsfonts}
\usepackage[pdftex]{color}
\begin{document}

\newcommand{\HII}{{\sc H\,i\,i}}
\newcommand{\Ha}{H$\alpha$}
\newcommand{\Haw}{\hbox{H$\alpha$\,$\lambda $6562.78}}
\newcommand{\ghafas}{{\sc GH$\alpha$FaS}}

\def\RR{$\mathfrak{R^2}$}
\def\kms{$\mbox{km s}^{-1}$}
\def\Myr{$\mbox{M}_\odot\mbox{ yr}^{-1}$}
\def\etal{et al.~}
\def\deg{^\circ}
\def\asim{\mathord{\sim}}
\def\farcs{\hbox{$.\!\!^{\prime\prime}$}}
\def\arcmin{\hbox{$.\!\!^{\prime}$}}

\title[Scalelength of disc galaxies]{Scalelength of disc galaxies}
\author[Fathi et al.]{Kambiz Fathi$^{1,2}$\thanks{E-mail: kambiz@astro.su.se}, Mark Allen$^{3}$,
Thomas Boch$^{3}$,
Evanthia Hatziminaoglou$^{4}$,\newauthor 
Reynier F. Peletier$^{5}$ \\ \ \\
$^1$Stockholm Observatory, Department of Astronomy, Stockholm University, AlbaNova Center, 106 91 Stockholm, Sweden\\
$^2$Oskar Klein Centre for Cosmoparticle Physics, Stockholm University, 106 91 Stockholm, Sweden\\
$^3$Observatoire de Strasbourg, UMR 7550 Strasbourg 67000, France\\
$^4$European Southern Observatory, Karl-Schwarzschild-Str. 2, 85748 Garching bei M\"unchen, Germany\\
$^5$Kapteyn Astronomical Institute, Postbus 800, 9700 AV Groningen, The Netherlands}

\date{Draft \today}
\pagerange{\pageref{firstpage}--\pageref{lastpage}} \pubyear{2010}
\maketitle

\label{firstpage}

\begin{abstract}
We have derived disk scale lengths for 30374 non-interacting disk galaxies in all five SDSS
bands. Virtual Observatory methods and tools were used to define, retrieve, and analyse the
images for this unprecedentedly large sample classified as disk/spiral galaxies in the LEDA
catalogue. Cross correlation of the SDSS sample with the LEDA catalogue allowed us to
investigate the variation of the scale lengths for different types of disk/spiral galaxies.
We further investigat asymmetry, concentration, and central velocity dispersion as indicators 
of morphological type, and are able to assess how the scale length varies with respect to
galaxy type. We note however, that the concentration and asymmetry parameters have to be
used with caution when investigating type dependence of structural parameters in galaxies.
Here, we present the scale length derivation method and numerous tests that we have carried
out to investigate the reliability of our results. The average $r$-band disk scale length
is $3.79$ kpc, with an RMS dispersion of $2.05$ kpc, and this is a typical value irrespective 
of passband and galaxy morphology, concentration, and asymmetry. The derived scale lengths presented here are representative for a typical galaxy mass of $10^{10.8\pm 0.54} \rm{~M}_\odot$, and the RMS dispersion is larger for more massive galaxies. Separating the derived scale lengths for different galaxy masses, the $r$-band scale length is $1.52 \pm 0.65$ kpc for galaxies with total stellar mass $10^{9}$--$10^{10}\rm{~M}_\odot$ and $5.73 \pm 1.94$ kpc for galaxies with total stellar mass between $10^{11}$ and $10^{12}\rm{~M}_\odot$. Distributions and typical 
trends of scale lengths have also been derived in all the other SDSS bands with linear 
relations that indicate the relation that connect scale lengths in one passband to another. Such
transformations could be used to test the results of forthcoming cosmological simulations of galaxy formation and evolution of the Hubble sequence.
\end{abstract}


\section{Introduction}
\label{sec:intro}
 
The exponential scale length of a galaxy disk is one of the most fundamental parameters to
determine its morphological structure as well as to model its dynamics, and the fact that
the light distributions are exponential makes it possible to constrain the formation
mechanisms \citep{Freeman70}. The scale length determines how the stars are distributed
throughout a disk, and can be used to derive its mass distribution, assuming a specific M/L
ratio. Ultimately, this mass distribution is the primary constraint for determining the
formation scenario \citep[e.g.,][ and referenced therein]{LP87,Dutton09}, which dictates
the galaxy's evolution. As the galaxy evolves, substructures such as bulges, pseudo-bulges,
bars, rings, and spiral arms may build up, and this will then considerably change the
morphology of the host disks \citep{CE93,Eetal05,Bournaud07}. The scale length value is
intimately connected to the circular velocity of the galaxy halo, which in turn relates
closely to the angular momentum of the halo in which the disk is formed \citep{DSS97,
Mo98}. Up to the last few years, cosmological simulations were limited to rather low resolution, were disks and spheroids were barely resolved, and generally limited to high redshifts, so reproducing realistic disk scale lengths for modern galaxies was clearly out of reach. The current simulations reach resolutions that allow resolving the disks from high redshift down to redshift zero, and subtle mechanisms changing the disk masses and scale lengths can be studied \citep[e.g., ][ listed in alphabetic order]{Ceverinoetal10,Governatoetal10,MartigBournaud10,Schayeetal10}, thus calling for a comprehensive observational determination of these parameters to test the state of the art cosmological simulations.

Previous observations of NGC, UGC, and low surface brightness galaxies have shown that
scale lengths span over a range of three orders of magnitudes \citep[e.g., ][]{Boroson81,
Romanishin83, vdKruit87, Schombertetal92, Knezek93, deJong96}. Any physical galaxy
formation scenario should be able to explain this wide range of values while simultaneously
explaining the similarities among disk galaxies throughout this range.

Analytic disk formation scenarios predict that, in cases where angular momentum is
conserved, the disk scale length is determined by the angular momentum profile of the
initial cloud  \citep{LP87}, and the scale length in a viscous disk is set by the interplay
between star formation and dynamical friction \citep{Silk01}. These processes form the
basis of a galaxy's gravitational potential, and determine the  
strength of gravitational
perturbations, the location of resonances in the disk, the formation and evolution of
spiral arms and bars, kinematically decoupled components in centres of galaxies, and the
dynamical feeding of circumnuclear starbursts and nuclear activity \citep[e.g.,][]{Eetal96,
Knapen04, Fathi04, KK04}.

Photometrically, one generally derived this scale length  
by azimuthally averaging profiles of the
surface brightness which is in turn decomposed into a central bulge and an exponential
disk, and when spatial resolution allows other components such as one or several bars and
rings can be taken into account. 

\begin{figure}
\begin{center}
  \includegraphics[width=.49\textwidth]{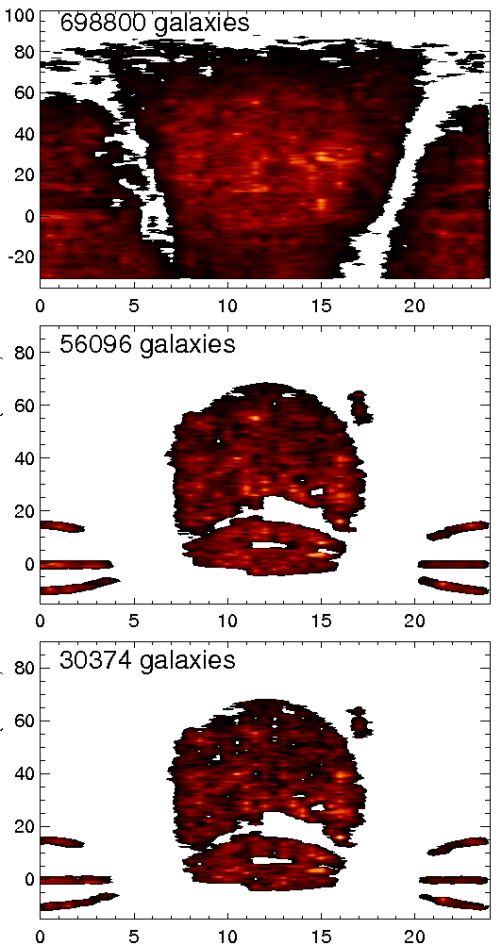}
  \caption{Top: right ascension and declination distribution of the galaxies for which the
LEDA services provide a Hubble classification number $\ge 1$ (irrespective of
classification error). Middle: same  for the 56096 SDSS galaxies fulfilling the first
sorting criteria described in section~\ref{sec:sample}, and Bottom: the
final sample of 30374 selected SDSS disk galaxies for which we have reliable 
scale $r$-band scale lengths (regardless of uncertainty in the morphological 
classification, see also section~\ref{sec:reliable}).} 
  \label{fig:radec}
\end{center}
\end{figure}
As images in different bands probe different optical depths and/or stellar populations, it
is likely that a derived scale length value should depend on waveband, and these effects
may vary as a function of galaxy type where different amounts of dust and star formation
are expected. Dusty disks are more opaque, resulting in 
larger scale length values in
bluer bands when compared with red and/or infrared images. Similar effects can also be
caused by differences in the stellar populations. 
Differences in scale length as a function of passband can therefore be used to
derive information about the stellar structure and contents of galactic disks.  
Both the effects of stellar populations and dust extinction have been subject to much
discussion over the years \citep[e.g.,][]{SdeV83, Kent85, Val90, vDetal95, Petal94,
Petal95, Betal96, Courteau96, Cunow98, BBA98, Petal01, Graham01, GdeB01, Cunow01, GH02,
MacA03, Cunow04, GW08}. A detailed and extensive analysis of the dust effects has also been
presented for a few tens of galaxies in \citet{Holwerda05} and subsequent papers by this
author, however, as noted by \citet{Petal94} and \citet{vDetal95}, the scale length alone
in different wavelengths in small sample cannot be used to break the age/metallicity and
dust degeneracies. Investigating the scale length variation as a function of inclination
for large numbers of galaxies is necessary to distinguish between the effects of
dust and stellar populations. 

The common denominator in all the previous studies is the roughly comparable sample sizes
(at most few hundred galaxies). Most studies have so far analysed individual galaxies, or
samples containing a few tens, and in unique cases a few hundred \citep[e.g.,][]{KvK91,
Courteau96}, galaxies. Although a number of great results from studies with the
Sloan Digital Sky Survey \citep[SDSS; ][]{Yorketal00} in the last years have appeared, 
these works have not studied the astrophysical parameters targeted here.

As a part of a European Virtual Observatory\footnote{http://www.euro-vo.org} Astronomical
Infrastructure for Data Access (Euro-VO AIDA) research initiative, we have undertaken a
comprehensive analysis of the scale length in disk galaxies using an unprecedentedly large
sample of disk galaxies. We have used the Virtual Observatory (VO) tools to retrieve data
in all ($u$, $g$, $r$, $i$, and $z$) bands from the sixth SDSS major data release
\citep[DR6; ][]{AMcetal08} which includes imaging catalogues, spectra, and redshifts freely
available. We use the LEDA\footnote{http://leda.univ-lyon1.fr} catalogue \citep{Petal03} to
retrieve morphological classification information about our sample galaxies, and those with
types defined as Sa or later are hereafter refereed to as disk galaxies (distribution of
both samples are presented in Fig.~\ref{fig:radec}).

In the present paper, we present the data retrieval and analysis method used to 
automatically derive the scale lengths for a sample of disk galaxies which contains 56096 objects
(described in section~\ref{sec:sdsssample}), and after rigourous tests described in
section~\ref{sec:scalelengths}, we find that a subset of 30374 of
these can be called reliable following these criteria. The scale lengths presented here relate only to the disk components, and we have tried to avoid the regions that could be dominated by the bulge component, in order to avoid complications related to the uncertainties of bulge-disk decomposition procedure \citep[as demonstrated in, e.g., ][]{KvK91}. In section~\ref{sec:results} we present the first results based on our unprecedentedly large sample of galaxies and finally discuss their implications in section~\ref{sec:discussion}.

\begin{figure*}
\begin{center}
  \includegraphics[width=.99\textwidth]{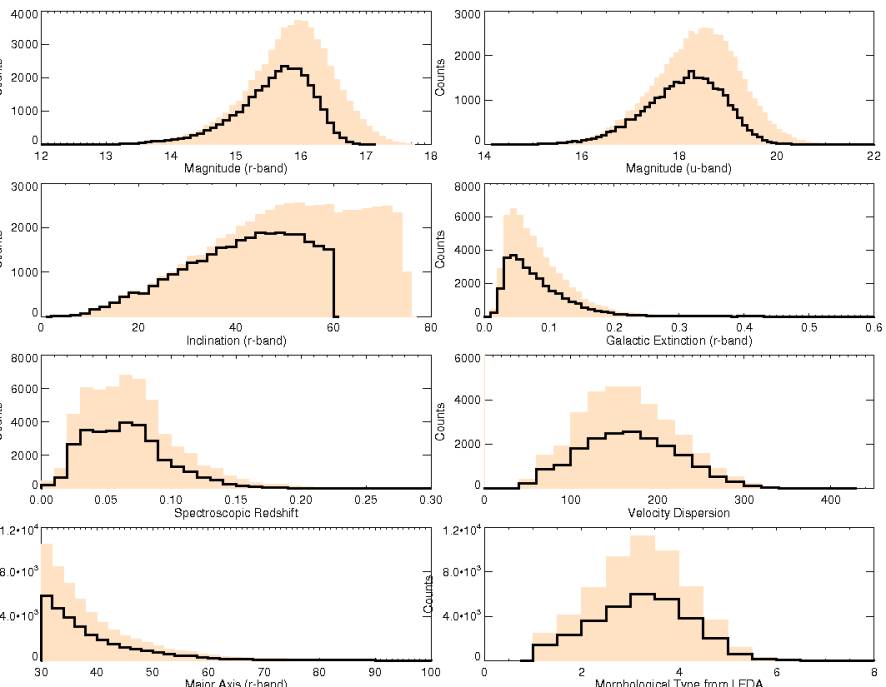}
  \caption{Distribution of some key parameters retrieved from the SDSS database and
morphological types from LEDA (bottom right). The filled histograms are the 56096 disk
galaxies described in section~\ref{sec:sample}, and the open histograms show the
distributions of the final 30374 galaxies for which we derive reliable scale lengths. The
distribution of the sample remains unchanged.} 
  \label{fig:design}
\end{center}
\end{figure*}

\section{Sample Galaxies from SDSS}
\label{sec:sdsssample}

\subsection{First Selection Criteria}
\label{sec:sample}
The DR6 provides imaging catalogues, spectra, and redshifts for the third and final data
release of SDSS-II, an extension of the original SDSS consisting of three sub projects: The
Legacy Survey, the Sloan Extension for Galactic Understanding and Exploration, and a
Supernova survey. The SDSS Catalogue Archive Server Jobs
System\footnote{http://casjobs.sdss.org} allow for a sample selection based on a number of
useful morphological and spectroscopic parameters provided for all objects. We use these
parameters and make a first selection of the entire SDSS DR6 sample. Various VO methods
were investigated to perform the download of the SDSS images, and the SkyView
\footnote{http://skyview.gsfc.nasa.gov} was chosen for this task. This service has the
advantage of being able to create fits cut outs centred at a given sky coordinate and with
a given size. Moreover, SkyView is able to re-scale the image backgrounds to the same
level, hence correcting for background level differences between the SDSS tiles. 

The image size is an important parameter to achieve a reliable sky subtraction which is necessary to derive realistic scale lengths, thus we require that the images cover an area at least three times the size of each galaxy. To optimise the data handling and keep low data transfer time from SkyView, we chose a constant image size of $900 \times 900$ pixels to be sampled for all galaxies, still being able to achieve a reliable sky subtraction. With these specifications, the image size is 3.2 MB with the typical download time of 16 seconds per image. This also includes the time that SkyView spends cutting, mosaicing, and re-scaling images. 

Our first selection criteria uses SDSS parameters to ensure that: 
\begin{description}
\item[1)] The object is a galaxy, and has good quality images available, i.e., quality keyword $\ge 2$.
\item[2)] The galaxy is at a position with low $r$-band Galactic extinction $A_r \le 1.0$. In reality, we find that 99\% of the sample have $A_r \le 0.25$.
\item[3)] For each galaxy SDSS provides spectroscopic redshift measurement, i.e., galaxy $r$-band magnitude $\le 17.7$.
\item[4)] The galaxy diameter is at least 60 pixels ($=24\arcsec$) and 
 at most 200 pixels ($=80\arcsec$ in diameter). The first criterion ensures that the
derived light profile samples the disk with at least 10 data points (for two-pixel wide rings) 
to derive the scale length, and the second criterion is for an optimised data retrieval procedure 
described above. Here we use the $r$-band isophotal semi major-axis $isoA$ and semi minor-axis $isoB$
as a measure for the galaxy size.
\item[5)] High inclination ($incl. \ge 70$ degrees) galaxies are removed to avoid selection
effect problems, but also since scale lengths for such systems are not reliable. The inclination is determined using the ratio between the semi minor axis $isoB$ and semi major axis $isoA$ in the $r$-band from the SDSS parameter list ($\cos i = isoB/isoA$).
\item[5)] No redshift cut was applied, however, Fig.~\ref{fig:design} shows that the sample extends out to redshift 0.3, with the typical redshift at $\log z = -1.2$ derived with $\ge 0.995$ confidence level, and with $15\%$ of the sample with $\log z \ge -1.0$. 
\end{description}

This first set of criteria leaves us with a total of 95735 galaxies. We use the LEDA
services to retrieve a numeric Hubble classification parameter $T$ for the galaxies in our
sample (more on this in section~\ref{sec:spirals}). We first download the entire LEDA
catalogue, which we cross-correlate with the SDSS sample using
TOPCAT\footnote{http://www.starlink.ac.uk/topcat} and only select the galaxies which, in
LEDA, are classified as spiral galaxies (i.e., $1 \le T \le 8$).
A total of 56096 Sa-Sd (i.e.,
$T$ between 1 and 8) spiral galaxies (see Fig.~1) were found, for which SDSS $u, g, r, i,
z$-band images were downloaded. In section~\ref{sec:spirals}, we further discuss whether
all these galaxies are well-classified disk or spiral galaxies.

In Fig.~\ref{fig:design}, we show the distribution of some key parameters retrieved from
the SDSS and LEDA database. This figure shows that the different sample selection stages do
not introduce any biases in our sample. It should be noted that, at this stage, we are
unable to determine whether the galaxies in our sample are isolated or disturbed systems,
as this information is not provided by any of the catalogues we have used. We make this
distinction using the asymmetry parameter described in \citet{Schadeetal95}.

\begin{figure}
\begin{center}
  \includegraphics[width=.49\textwidth]{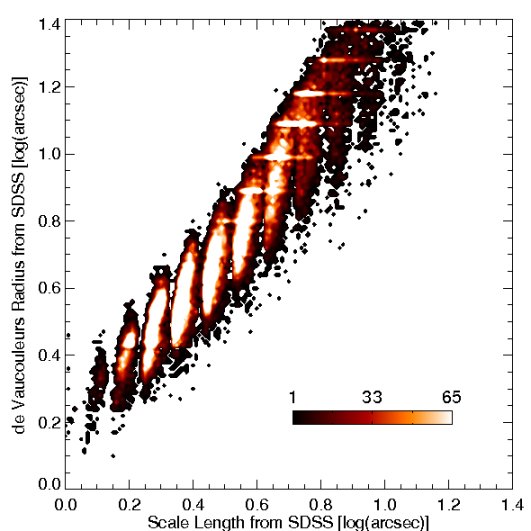}
  \caption{Density plot of the de Vaucouleurs effective radius ($y$-axis) versus exponential disk radius ($x$-axis) provided by the SDSS service for the entire disk galaxy sample. The odd clustering of the data (overdensities around discrete values) show the strong systematic effects in these two parameters provided by the SDSS team.} 
  \label{fig:rddev}
\end{center}
\end{figure}
\subsection{Scale lengths from SDSS}
\label{sec:rddev}
The first issue that arises at this point is the fact that SDSS services provides users
with the disk scale length as well as de Vaucouleurs effective radius for each galaxy (in
all bands), and that, in principle, these values could be used to carry out our analysis. In
Fig.~\ref{fig:rddev}, we show that the values provided by the SDSS services show anomalies
that are beyond our satisfaction for carrying out our analysis. The plot shows peculiar
systematic effects in their derivation of the de Vaucouleurs radii and scale lengths around
some discrete values marked by the overdensities, the source and explanations for which we
cannot find. We thus decide to re-calculate the scale lengths.

\begin{figure*}
\begin{center}
  \includegraphics[width=.95\textwidth]{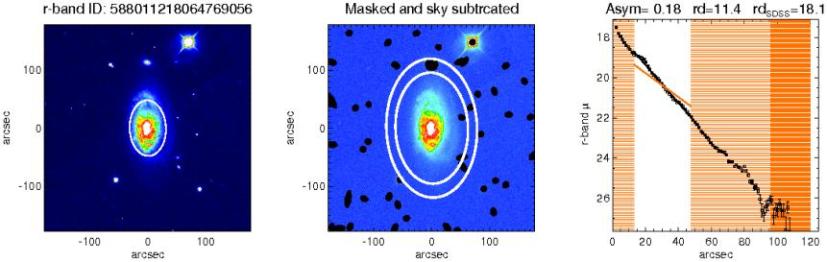}
  \includegraphics[width=.95\textwidth]{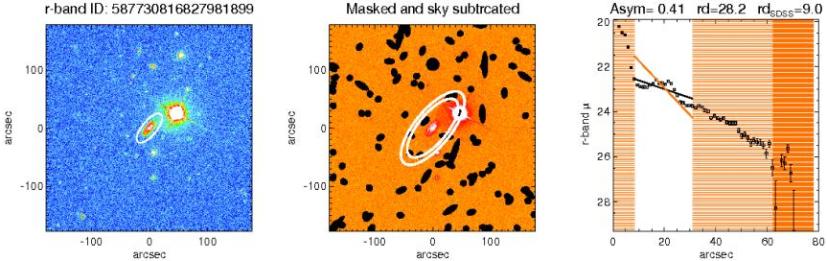}
  \includegraphics[width=.95\textwidth]{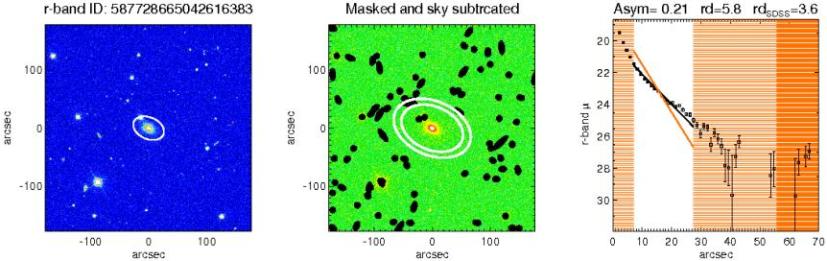}
  \caption{Three randomly selected galaxies for which we illustrate the procedure for deriving the scale length (see section~\ref{sec:scalelengths}). For each galaxy, the left panel shows the $r-band$ image and the ellipse with axis-ratio $isoB/isoA$. The middle panel shows the sky-subtracted image, the sky region $2.0\pm 0.25 \times isoA$ outlined by two corresponding ellipses, and the SExtractor sources masked. The right panel shows the light profile using the zero point from the SDSS, and linear fit to the disk region as described in section~\ref{sec:scalelengths}. At the top this panel, the asymmetry parameter, our derived disk scale length (black linear fit) and the scale length from SDSS (arbitrarily shifted red line) are stated.} 
  \label{fig:thecode}
\end{center}
\end{figure*}

\section{Deriving Scale lengths and Asymmetries}
\label{sec:scalelengths}
\subsection{Using IDL and GDL}
To derive the disk scale length, we use some important parameters provided by the SDSS in
order to constrain galaxy geometry as well as the location of the sky region. These are
semi minor axis $isoB$, semi major axis $isoA$, isophotal position angle $isoPhi$, and for
consistency, we use these $r$-band  quantities in also in all other bands.  
Our scale length
derivation routine uses standard IDL routines, though due to license limitations this code
can only be executed once, and hence is estimated to take a long time to run. Using one
single IDL session, we would need 47 days to derive the scale lengths for the entire
sample, thus in order to speed up the process, we decided to run this computation on a
cluster of machines located at Centre de Donn\'ees astronomiques de Strasbourg CDS. Since
the freely available IDL virtual machine does not allow one to launch batch queries, and
since installing an IDL licence on each cluster node was not an option, we used the open
source clone of IDL, GNU Data Language GDL \citep{GDL}. We found out that a few IDL
functions were either missing or behaving improperly, thus requiring minor tweaking in our
code. Then, we ran both IDL and GDL code on the same subset of SDSS images, in order to
check the reliability of the GDL output.

Once the GDL code was installed on each node of the cluster, 
the 56096 SDSS images were 
put to an iRods\footnote{iRods (www.irods.org) is a distributed data management system,
which provides a distributed storage environment to easily store and share files}
installation deployed at CDS. Finally, the scale length computation  was  
launched on
the cluster, using the following architecture:

\begin{enumerate}
\item A Java program holds the list of galaxies to process, and - for each object of this list - sends a message to the cluster, asking to spawn a new job (i.e., launch the corresponding computation).
\item The cluster master node receives the request, and dispatches it to the cluster node with the smallest CPU load.
\item The cluster node then downloads from iRods the $u,g,r,i,z$ images corresponding to the galaxy to process, runs the GDL code, and sends back the computed result to iRods.
\item If the computation fails, it will be re-sent to another node until success
\end{enumerate}

The CDS cluster has proven to be very stable and reliable, though some problems were found
in the dispatching algorithm, resulting sometimes in overloading some of the nodes while
some others were idle. Four nodes of the cluster were dedicated to our computation. As the
total computation time is roughly proportional to the number of involved nodes, allocating
ten times more nodes would have decreased this time by a factor ten. 
This would only be true if the
computation service were to be close  to the data, 
so that the transfer time  would be negligible with
respect to the computation time. In theory, on the CDS cluster, with four dedicated nodes,
we should have been able to process 11200 galaxies per day, 
however in reality, the cluster only
processed between 8500 and 9000 galaxies per day, which could be explained by some
inefficiency in the dispatching algorithm. To conclude, 
using the CDS, we have been able to
derive the scale lengths for all 56090 galaxies in all five SDSS bands in less than a week.

\subsection{The Procedure}
The procedure to derive scale length and calculate asymmetry parameters from the SDSS images is illustrated in Fig.~\ref{fig:thecode} and carries out the following steps:
\begin{itemize}
\item Reading the image and assigning the pre-determined $r$-band parameters from a file
that contains all SDSS parameters for the entire sample.

\item Selecting the sky region as the ellipse encompassing the range $2.0\pm 0.25 \times
isoA$. This is marked as a darker shaded region in the rightmost panels in
Fig.~\ref{fig:thecode}. The mean value of this region, using Tukey's bi-weight mean
formalism described in \citet{MT77}, is used to calculate the sky level for sky subtraction
as well as setting the background level.

\item To remove foreground 
stars and point sources from the image, we extract point sources
with SExtractor \citep{BA96}, by selecting all point sources that are larger than 4 pixels
in size and more than 3 $\sigma$ above the background level. All pixels belonging to these
sources are then masked out, and we note that our selection could include bright
star-forming regions and small background galaxies in these sources.

\item Using the asymmetry parameter definition of \citet{Schadeetal95} and \citet{Conselice03}, we calculate the asymmetry parameter 
\begin{equation}
\label{eq:a}
A=\sum \|I-I_{180}\|/ \sum I,
\end{equation}
where $I$ is the sky subtracted galaxy image intensity and $I_{180}$ is that for the image rotated by 180 degrees around the galaxy centre. It should be noted here that the asymmetry criterion applied here removes ongoing mergers and galaxies with companions at a projected distance of about twice the galaxy radius, and here we take the results of \citet{Conselice03} at face value, that $A \ge 0.35$ means that the system is disturbed.

\item Using the $isoB$, $isoA$, $isoPhi$ parameters from SDSS, we then section each galaxy into 2-pixels wide ellipses oriented at the major axis position angle $isoPhi$ and with minor-to-major axis ratio $b/a=isoB/isoA$. The bi-weighted mean surface brightness value within each ellipse is calculated to compile the galactocentric surface brightness profile $\mu(r)$ for each galaxy.
\end{itemize}

\begin{figure}
\begin{center}
  \includegraphics[width=.49\textwidth]{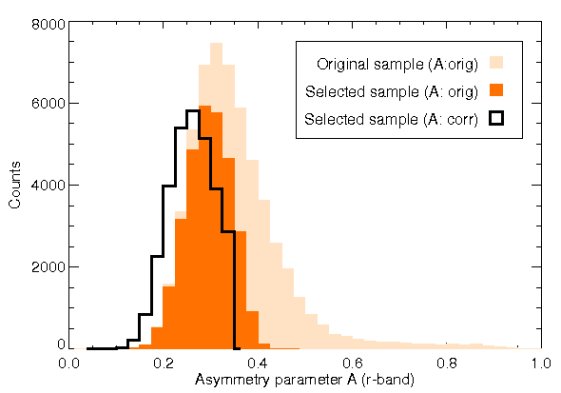}
  \caption{Asymmetry parameter for all 56096 disk galaxies assuming the objects are located
in the centre of the image (grey histogram), and when the galaxy image is rotated around
the ``true'' galaxy centre found by centroid fitting (red histogram). The final set of
30374 low-inclination and low-asymmetry galaxies for which we derive the scale lengths are
shown with the black histogram.} 
  \label{fig:a}
\end{center}
\end{figure}

In spatially resolved systems, surface brightness profiles are commonly fitted by a multiple of parametric functions in order to describe the contribution of different components to the observed profile. A de Vaucouleurs \citep[$r^{1/4}$, ][]{deV48} or S\'ersic \citep[$r^{1/n}$, ][]{Sersic68} law is typically used for the innermost part of the disk, and for the outer parts an exponential function of the form
\begin{equation}
\label{eq:exponential}
\mu(r) = \mu_0 + 1.086 \,\frac{r}{r_d},
\end{equation}
is used, where $\mu_0$ is the central surface brightness, $r$ is the galactocentric radius,
and $r_d$ is the disk scale length of the outer disk. 
In addition to these two components, other functions
may be used to fit the halo component, bars, rings, and other structures in the galaxies
\citep[e.g.,][]{Petal01}, and the fits can be applied to one-dimensional light profiles or
directly on two-dimensional images \citep{BF95}. Here we fit
equation~(\ref{eq:exponential}) to the one-dimensional surface brightness profiles.

Running the fully automated fitting algorithm on all retrieved images, we found a
number of artefacts which cause problems for applying the code successfully. These include:
\begin{itemize}
\item SkyView does not deliver the image for the galaxy in all bands, i.e., a blank image
has been transferred and stored. The first query delivered 892 blank images, and a second query on the blank images delivered successfully less than 1\% of the images.
\item The galaxy is positioned such that there  are no adjacent tiles 
observed yet, and thus  a
large part of the retrieved image is filled by SkyView  with blank pixels.
\item The galaxy is too faint in a given band to deliver reliable surface brightness profile, i.e., the linear fit results in a negative slope. 
\item Man-made satellites passing too close to the galaxy position.
\end{itemize}

\subsection{Reliable Scale Lengths}
\label{sec:reliable}
Saturated stars near the objects cannot be masked properly using SExtractor (due to
undetermined source radii). Moreover, strong galaxy interactions and noisy images
introduce errors in the derived scale lengths. We selected 
randomly a few hundred images for
which we plot the surface brightness profiles with corresponding linear fits. Visual
inspection  showed  
that the routine runs as expected. Saturated stars, if far away from a
galaxy (further then $2.25\times isoA$, i.e., the outermost sky pixel) do not introduce any
errors in the derived parameters as they are not considered at any stage. If close to a
galaxy, they can be regarded as interactions. Interactions between galaxies can be
quantified following \citet{Conselice03} and  \citet{Conseliceetal03} who found that
interacting or merging galaxies  mostly  have asymmetry parameter $A\ge 0.35$. 

When applying equation~(\ref{eq:a}), it is crucial to rotate the images around the centre of the galaxy, as minor offsets can significantly overestimate $A$. We apply a centroid fitting to our sample, and find that the images generated by SkyView are off-centred by about half a pixel. This offset, although minor in terms of pixels and arcseconds, significantly over-estimates $A$ for our galaxies (see Fig.~\ref{fig:a}).

\begin{figure}
\begin{center}
  \includegraphics[width=.49\textwidth]{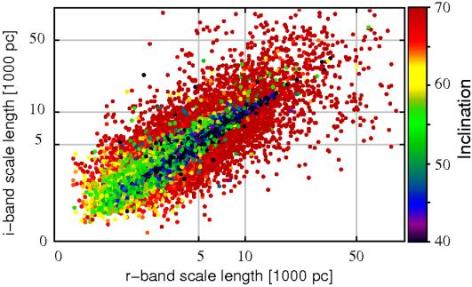}
  \caption{Scale lengths form the $i$-band images ($y$-axis) versus those from $r$-band
images ($x$-axis), where the inclination of the galaxy is shown by the colour bar on the
right of the figure. Although the points fall on a 1:1 slope, the galaxies with $incl. <
60$ degrees have smaller scatter ($\mathfrak{R^2} > 0.90$, see text in
section~\ref{sec:reliable}).} 
  \label{fig:incl}
\end{center}
\end{figure}

If  the galaxy image is deep enough, it is expected that the 
scale length  values  in
two adjacent bands should be similar. As images in all SDSS bands are not equally deep,
we investigate the $r$ and $i$-band images for this purpose only, since these two filters
are comparable, and sufficiently adjacent to deliver almost identical scale lengths.
Plotting the corresponding scale lengths, shown in Fig.~\ref{fig:incl}, we find that
galaxies with high inclination $incl. \ge 60$ degrees) are the objects that introduce the
large scatter in this diagram. 

We use Pearson's product moment correlation coefficient to calculate the coefficient of determination \RR\ according to the standard formula 
\[\mathfrak{R^2} = 1-\frac{\displaystyle \sum^N_{j=1} (X_j-\hat{X}_j)^2}{\displaystyle \sum^N_{j=1} (X_j-\overline{X})^2},\]
where $N$ is the number of data points, $X_j$ are the measured data, $\hat{X}_j$ are the estimated values given by linear regression, and $\overline{X}$ is the mean value of the measured data points. In simple statistical terms, the numerator is termed the total sum of squares, and the denominator is the error sum of squares, and the coefficient \RR\ provides the percent of the variation that can be explained by the linear regression equation, and therefore is a useful measure for the variance of one variable that is predictable from the other variable. If the regression line passes exactly through 50\% of the data points, it would be able to explain half of the variation of the linear fit, and would result in \RR=0.5. Throughout our analysis, we trust a correlation if $\mathfrak{R^2}\ge 0.9$, and $\mathfrak{R^2}\le 0.68$ is considered insignificant (c.f., less than one-sigma confidence level is insignificant).

The dispersion of the data points around the 1:1 line can is then measured by calculating \RR\ for different inclination bins, and we find that this parameter remains above 0.90 for $incl. \le 60$, hence keep all the galaxies with $incl. \le 60$ degrees. We will later find that combining the inclination and asymmetry restriction will deliver even higher coefficient of determination between the scale lengths in different bands.

Finally, we find that for five galaxies, the SDSS spectroscopic redshifts are larger than 1, whereas the rest of the sample has redshift $<0.3$. Despite the small redshift errors, we find the redshifts for these five galaxies unrealistic and we choose to remove them from our sample.

{\em Thus, applying these cuts, we derive scale lengths for 30374 disk galaxies that we
argue are reliable, given the arguments mentioned above.}

\begin{table*}
\caption{Disk scale length and central surface brightness for different assumed disk range as a fraction of $isoA$ as illustrated in Fig.~\ref{fig:diskrange}. The values presented here have been derived for a random sub sample of 800 galaxies with formal errors given in brackets, and all values are normalised to the scale length derived in the range 25\%--100\%.}
\centering \tabcolsep=6pt
\begin{tabular}{lcccc}
\hline \hline
Fitted $isoA$ Range & ($r$-band) $r_d \over r_d (25\%-100\%)$ & ($r$-band) $\mu_0 \over \mu_0 (25\%-100\%)$ & ($i$-band) $r_d \over r_d (25\%-100\%)$ & ($i$-band) $\mu_0 \over \mu_0 (25\%-100\%)$\\ \hline 
 0\%--100\% & 0.91(0.19) & 0.87(0.18) & 1.07(0.12) & 1.09(0.12)\\
20\%--100\% & 0.98(0.08) & 0.97(0.07) & 1.02(0.05) & 1.03(0.05)\\
25\%--100\% & 1          & 1          & 1          & 1         \\
30\%--100\% & 1.07(0.15) & 1.09(0.18) & 0.97(0.07) & 0.96(0.08)\\
30\%--100\% & 1.14(0.27) & 1.12(0.17) & 0.93(0.12) & 0.93(0.09)\\
40\%--120\% & 1.20(0.28) & 1.25(0.29) & 0.87(0.16) & 0.87(0.16)\\ \hline
\end{tabular}
\label{tab:diskrange}
\end{table*}
\begin{table*}
\caption{Disk scale length and central surface brightness for different assumed sky range as a fraction of $isoA$ as illustrated in Fig.~\ref{fig:skyrange}. The values presented here have been derived for a random sub sample of 800 galaxies, normalised to the sky range $2.00\pm 0.25$, with formal errors given in brackets.}
\centering \tabcolsep=9pt
\begin{tabular}{lcccc}
\hline \hline
Fitted Sky Range & ($r$-band) $r_d \over r_d (2.0\pm 0.25)$ & ($r$-band) $\mu_0 \over \mu_0 (2.0\pm 0.25)$ & ($i$-band) $r_d \over r_d (2.0\pm 0.25)$ & ($i$-band) $\mu_0 \over \mu_0 (2.0\pm 0.25)$\\ \hline 
$1.75\pm 0.25$ & 0.99(0.03) & 0.99(0.03) & 1.00(0.01) & 1.00(0.01)\\
$2.00\pm 0.25$ & 1          & 1          & 1          & 1         \\
$2.25\pm 0.25$ & 1.00(0.03) & 1.01(0.04) & 1.00(0.01) & 1.00(0.02)\\
$2.50\pm 0.25$ & 1.01(0.05) & 1.01(0.07) & 1.00(0.02) & 1.00(0.03)\\
$2.75\pm 0.25$ & 1.01(0.06) & 1.01(0.09) & 1.00(0.03) & 1.00(0.04)\\ \hline
 \end{tabular}
\label{tab:diskrange}
\end{table*}
\begin{figure}
\begin{center}
  \includegraphics[width=.49\textwidth]{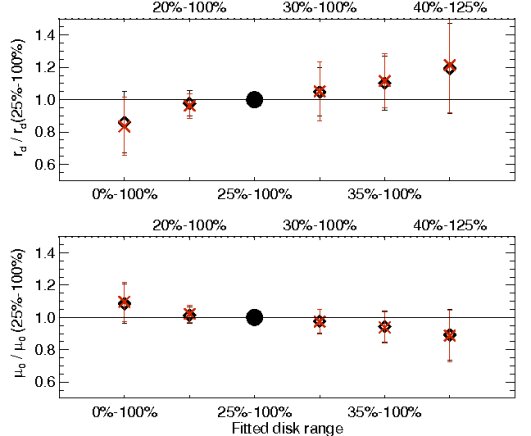}
  \caption{Disk scale length and central surface brightness for different assumed disk range as a fraction of $isoA$ parameter provided by the SDSS. The values illustrated here have been derived for a random sub sample of 800 galaxies in $r$-band (black diamonds) and $i$-band (red crosses). All values have been normalised to that illustrated by the black circle, which is the disk range we assume throughout this work.} 
  \label{fig:diskrange}
\end{center}
\end{figure}
\begin{figure}
\begin{center}
  \includegraphics[width=.49\textwidth]{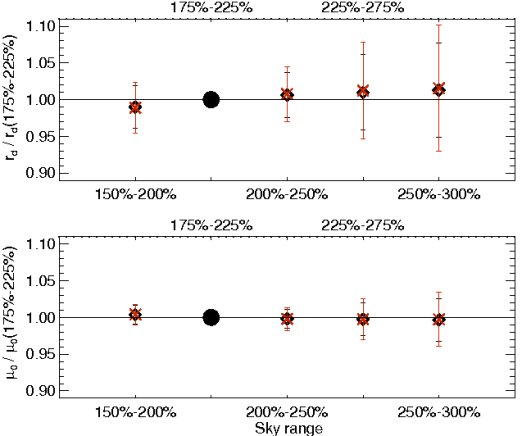}
  \caption{Disk scale length and central surface brightness for different assumed sky range as a fraction of $isoA$ parameter provided by the SDSS. The values illustrated here have been derived for a random sub sample of 800 galaxies in $r$-band (black diamonds) and $i$-band (red crosses). The values have been normalised to that illustrated by the black circle, which is the sky range we assume throughout this work.} 
  \label{fig:skyrange}
\end{center}
\end{figure}
\subsection{Disk and Sky Ranges}
\label{sec:diskranges}
Here, we will not delve into the intricacies of fitting the light curves, but focus on the
determination of the scale length of the exponential disk. Despite the long tradition (see
references in section~1), the important and comprehensive study by \citet{KvK91} showed
that the errors in these data are still significantly large ($\approx 25\%$), especially if
they were obtained from photographic plates. The uncertainties depend on image depth, image
sky coverage, data reduction, disk region fitting, the order in which bulges, bars, or
other components are fitted. These matters become more complicated when analysing with SDSS
images which are relatively shallow, and even more so when automatically fitting thousands
of galaxies which cover a wide range of brightness and morphologies. To avoid complications
that are not related to the nature of our analysis \citep[e.g., ][]{Fathi03}, we have
decided to derive the disk scale length simply by fitting an exponential profile to a
pre-defined disk region of each galaxy, i.e., the region where we assume the light to be
dominated by the exponential profile. This  means that we are simply cutting out the
central regions of the galaxies where bulges and strong bars are expected.

We determine the disk region by empirically fitting the equation~(\ref{eq:exponential}) to
a set of ranges where we expect the disk to dominate the derived surface brightness
profiles. We use the $isoA$ parameter to estimate this range, and randomly select 800
galaxies, to which we apply this test both in $r$ and $i$-band.  In
Fig.~\ref{fig:diskrange}, we show the resulting disk scale lengths when fitting the regions
presented in Table~\ref{tab:diskrange} , and when normalised to our nominal 25\%--100\% of
the $isoA$ radius, we find that the derived scale lengths change by less than 10\% for a
wide range of assumed disk ranges (seen as the unshaded region in the rightmost panels in
Fig.~\ref{fig:thecode}). We further note that the distribution of the data points for each
test, normalised to the 25\%--100\% $isoA$ range, is well represented by a Gaussian, and
the error bars in Fig.~\ref{fig:diskrange} are indeed symmetric.

For a similar test, we assume sky regions at different distances from each
galaxy centre and assess the effect of the sky subtraction on the derived scale 
lengths. Applied to the same randomly selected 800 galaxies, we found that, 
assuming that the sky is represented by the $2.0\pm 0.25 \times isoA$ region, 
robust scale length and surface brightness measurements are delivered.

\begin{figure}
\begin{center}
 \includegraphics[width=.49\textwidth]{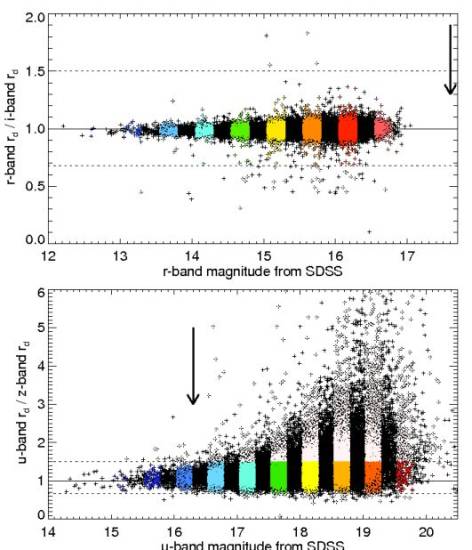}
  \caption{Scale length ratio versus magnitude for two pairs from table~\ref{tab:ugriz}.
Bins of 0.2 magnitude are used to scan the data points "from left to right", and when less
than 95\% of the ratios are inside  the dotted lines, 
that magnitude limit is taken to
be the faintest magnitude where we trust the scale lengths for these two bands. Here we
show the best case $r, i$-pair (top) and the worst case $u, z$-pair (bottom). In each
panel, the arrow indicates the cutting limit presented in table~\ref{tab:ugriz}.} 
\label{fig:magcut}
\end{center}
\end{figure}
\begin{table}
\caption{Magnitude cuts applied to the final sample of 30374 galaxies as described in 
section~\ref{sec:scalelengths}. To apply the cut to each pair, a plot similar to
Fig.~\ref{fig:magcut} was set up, and the magnitude cut was decided accordingly.}
\centering \tabcolsep=5pt
\begin{tabular}{ccr}
\hline \hline
Filter Pair & Upper Magnitudes & Number of Galaxies\\ \hline
$g$ and $r$ & $g<17.70$ \ \ \&\ \  $r<17.70$ & 30201\\
$g$ and $i$ & $g<19.95$ \ \ \&\ \  $i<16.65$ & 30371\\ 
$g$ and $z$ & $g<17.70$ \ \ \&\ \  $z<15.53$ & 27319\\ 
$r$ and $i$ & $r<17.70$ \ \ \&\ \  $i<16.65$ & 30374\\
$r$ and $z$ & $r<17.70$ \ \ \&\ \  $z<15.53$ & 27329\\
$i$ and $z$ & $i<15.89$ \ \ \&\ \  $z<15.53$ & 27264\\ \hline
$u$ and $g$ & $u<16.79$ \ \ \&\ \  $g<14.94$ & 847\\
$u$ and $r$ & $u<16.54$ \ \ \&\ \  $r<13.56$ & 132\\
$u$ and $i$ & $u<16.79$ \ \ \&\ \  $i<13.14$ & 123\\
$u$ and $z$ & $u<16.29$ \ \ \&\ \  $z<12.78$ & 88\\\hline

\end{tabular}
\label{tab:ugriz}
\end{table}
\subsection{Scale Lengths in $u, g, r, i, z$-bands}
\label{sec:ugriz}
Although the SDSS is one of the most influential and ambitious astronomical surveys, the
depth of its images in all bands are not equal. Here we have chosen to analyse only the
galaxies for which SDSS provides spectroscopic redshifts (in order to investigate the
redshift evolution the parameters we derived), where SDSS is complete for $r$-band
magnitude $<17.7$. The images in other bands are not equally deep and/or complete to this
magnitude limit, partly due to the significantly different transmission curves for the
different filters. Including atmospheric extinction and detector efficiency, the peak
quantum efficiency of the system in $u$ and $z$-bands are $\approx 10\%$, $g$ and $i$-band
$\approx 35\%$, and $r$-band $\approx 50\%$. Thus it is necessary to apply a magnitude cut
which varies depending on the band, fainter than which we are not able to derive reliable
scale lengths. 

For each pair of SDSS filters, we expect that the scale length variation larger than a
factor 1.5 is unphysical. We determine the magnitude limit for a pair of filters by
plotting the scale length ratio versus magnitude in one of the bands (see
Fig.~\ref{fig:magcut}), and scan the values from the brighter to the fainter levels in
fixed bins of 0.2 magnitudes. Once we reach a magnitude where less than 95\% of the scale
length ratios is smaller than 0.67 or larger than 1.5  
(i.e., above or below the horizontal dotted lines in Fig.~\ref{fig:magcut}), we stop the
scan and select this value for the faintest magnitude level in that band for which we trust
the scale lengths. As shown in Fig.~\ref{fig:magcut}, this procedure very clearly
demonstrates the noise in different bands, and how the values presented in
table~\ref{tab:ugriz} have been established. In the given examples, the scale lengths in
$r$-band when compared to $i$-band are complete to an $r$-band magnitude of 17.70
(indicated by the arrow), and the $u$ versus $z$-band is complete to an $u$-band magnitude
of 16.29 (indicated by the arrow).

\begin{figure}
\begin{center}
 \includegraphics[width=.49\textwidth]{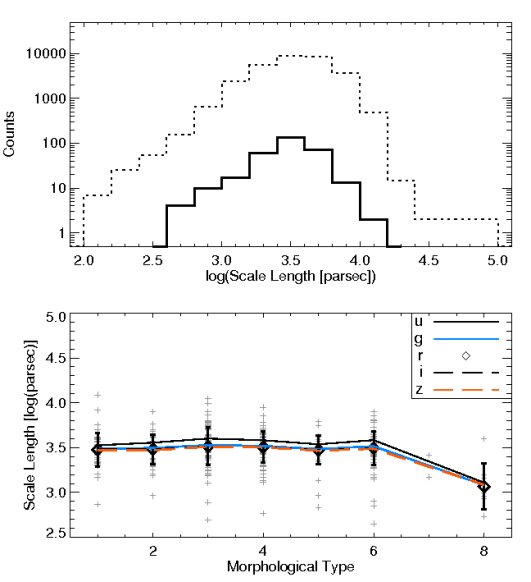}
  \caption{Top: Distribution of the reliably derived $r$-band scale lengths for the entire sample of 30374 galaxies (dotted histogram) and the 309 morphologically well-classified galaxies (solid histogram). Bottom: Scale length versus morphological type for the 309 galaxies which have been morphologically classified accurately. The $r$-band has been used with the scale length in $u,g,i,z$ bands i plotted in the middle panel. The error bars for all bands are comparable, and here we only show these for the $r$-band values.} 
\label{fig:typeArd}
\end{center}
\end{figure}

\begin{figure}
\begin{center}
 \includegraphics[width=.49\textwidth]{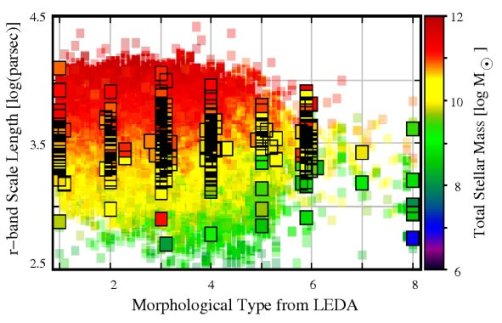}
  \caption{Scale length versus morphological type for the entire sample of 30374 galaxies, with the 309 well-classified galaxies marked with black squares. The colour represents the total stellar mass for each galaxy.} 
\label{fig:rdtypemass}
\end{center}
\end{figure}
\section{Results}
\label{sec:results}
\subsection{Scale Length versus Morphology}
\label{sec:spirals}
The morphological classification scheme of \citet{Sandage61}, is designed based on visual
inspection of basic features of galaxies which relates them to their formation and
evolution histories. While this classification scheme is somewhat subjective, in the past
years, numerous efforts have been made to define quantitative versions of this
classification scheme \citep[e.g., ][]{BF92,Doietal93,Abrahametal94,Yamauchietal05}. 

The numeric morphological types presented in the LEDA catalogue are a compilation of the
morphological types encoded in the de Vaucouleurs scale as well as the luminosity class
(van den Bergh's definition). There is also information about the presence of bars and
rings, but we do not consider these for the present paper mostly since this information is
only available for minor fraction of the sample. More details about the classification of
galaxies can be found in the Level 5 of the NASA/IPAC Extra-galactic Database (NED). The
morphological types of LEDA have been compiled using from \citet{VV63, N73, L82, deV91};
and \citet{L96}. 

We select only the galaxies which are classified using the numeric type = 1 (i.e., Sa) up
to and including numeric type = 8 (i.e., Sdm). Most of the values presented in
Fig.~\ref{fig:design} are subject to errors larger than 1, 
typically smaller for fainter galaxies,
but they seem not to depend much on other parameters such as asymmetry, redshift, etc. To
analyse the dependency of the parameters with respect to morphological type, we strictly
only use the galaxies for which the morphological type error is smaller than 0.5. These are
309 galaxies from our final sample of 30374 galaxies, for which we investigate how the
scale length and asymmetry parameter depends on morphology.

Typically, scale lengths for disk galaxies are not expected to depend on Hubble
morphological type (\citet{deJong96,GdeB01}) for types ranging between 1 and 6. 
Here, we analyse our derived values in this
context first by only using the galaxies for which we only find morphological
classifications with corresponding errors smaller than 0.5, i.e., the 309 galaxies
explained in section~\ref{sec:spirals}. In Fig.~\ref{fig:typeArd}, we plot the $r$-band
scale length and morphological types, and find that our sample is fully
consistent with previous results showing that the absolute value of the scale 
length is independent of type. We transform the
scale length to parsec units by using the spectroscopic redshifts provided by the SDSS and
ignore local flows. Our scale length values agree with those derived by previous authors
\citep[e.g., ][]{vdKruit87, Cunow01, deJong96}; we find that the average $r-band$
scale length for the entire sample is $3.79 \pm 2.05$ kpc, and that for the 309 galaxies
with reliable morphology is $3.3\pm 1.6$ kpc (see top panel of Fig.~\ref{fig:typeArd}).
Further discussion is provided in section~\ref{sec:wavelength}, and the errors are root mean square (RMS) values.

In combination with the mass determination described in section~\ref{sec:mass}, we find that the mass does play a certain role in the behaviour of Fig.~\ref{fig:typeArd}. Although out to type $T=6$ scale length are constant, the later type galaxies ($T>6$) are generally those of lower mass, and hence in agreement with Fig.~\ref{fig:rd_mass}. These lower mass galaxies have smaller scale lengths, thus the scale length decrease for late types is intimately linked with the lower left corner of Fig.~\ref{fig:rd_mass}. However, it should be noted that here we only have used the galaxies with robust morphological classification, and have a smaller number of low-mass galaxies as compared with high-mass galaxies (9 galaxies with total stellar mass $10^8-10^9 \rm{~M}_\odot$, 40 galaxies with total stellar mass $10^9-10^{10} \rm{~M}_\odot$, and 207 galaxies with total stellar mass $10^{10}-10^{11} \rm{~M}_\odot$).

\begin{figure}
\begin{center}
 \includegraphics[width=.49\textwidth]{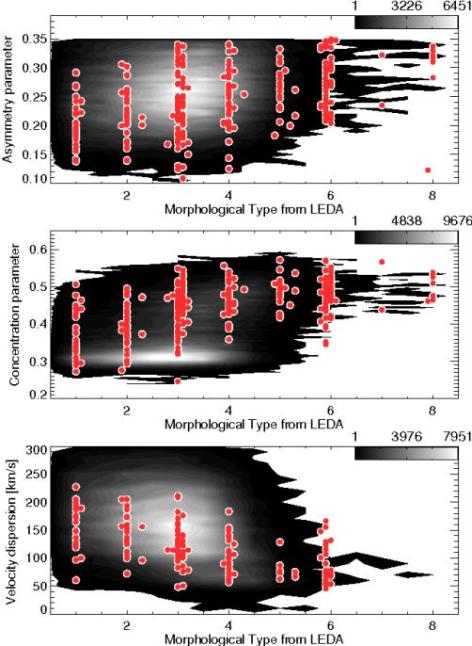}
  \caption{Asymmetry, concentration, and velocity dispersion as type indicator for the full sample (grey density plot) and for accurately (type error $\le 0.5$) classified galaxies in LEDA.} 
\label{fig:typeCASigma}
\end{center}
\end{figure}
\begin{figure}
\begin{center}
 \includegraphics[width=.49\textwidth]{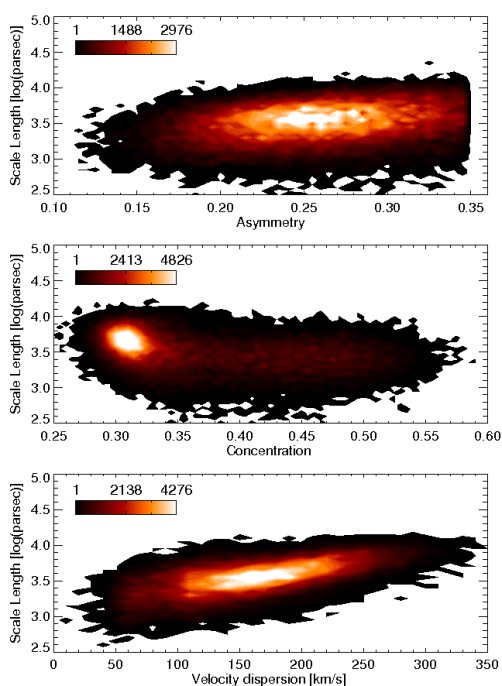}
  \caption{Scale length versus asymmetry and concentration parameter and stellar velocity dispersion for the full sample of 30374 galaxies.} 
\label{fig:CAsigma}
\end{center}
\end{figure}

We now cross-correlate our sample with the morphologically classified bright
galaxy catalogue of \citet{Fuku07}. Their catalogue contains 2275 galaxies classified by
visual inspection of SDSS images in the $g$-band. We find 283 objects overlapping between
the two samples. The small overlap is partly due to the fact that around half of the objects in \citet{Fuku07} are early-type galaxies (RC3 type $T<1$), and partly since 
it is essentially the overlap between the LEDA sample and that of \citet{Fuku07}. Moreover, our sample has an upper limit for the galaxy sizes due to our preferred strategy for using SkyView (see section~\ref{sec:sample}). From this sample of 283 galaxies, only 45 galaxies have been accurately classified (their T error $\le 1$) in \citet{Fuku07}. For these
objects, we find weak correlation between the morphological classification from LEDA
and those from \citet{Fuku07}. 

In a similar fashion to \citet{Shimasaku01}, we define the {\em "inverse"} concentration parameter as the
ratio between the radii containing 50\% and 90\% of the Petrosian flux respectively,
$r_{50}/r_{90}$ provided by the SDSS services. As a consistency check, we ensure that we
reproduce Fig.~10 of \citet{Shimasaku01}, i.e., that morphological classification provided
by LEDA correlates with concentration parameter. Furthermore, we find that although the
vast majority of our sample have very large morphological classification uncertainties (T error $\ge 1$), the
concentration parameters that we calculate for all 30374 galaxies indicate that, in
agreement with \citet{Shimasaku01}, they are disk galaxies. 

Regarding the correlation of the concentration parameter with morphological type, for the
309 well-classified galaxies $\mathfrak{R^2} = 0.31$, and for the full sample
$\mathfrak{R^2} = 0.16$. Although we find these values unconvincing as firm correlations,
we acknowledge a clear trend that concentration parameter is increasing with galaxy
morphological type. Likewise, the spread of points in asymmetry-type and velocity
dispersion-type diagrams are very large, and the coefficients of determination even lower
than that of the concentration parameter, however, here also the trend is acknowledged.

Given that robust morphological classification is known only for a very small subset of our
entire sample, we invoke other parameters in order to be able to further investigate the
scale lengths for the full sample. Following the above arguments, and their consistency
with the previous findings by, e.g., \citet{Conselice97, Shimasaku01}, and \citet{Fuku07}
we find it instructive to invoke these parameters as type indicators (see
Fig.~\ref{fig:typeCASigma}). For example, we find that the asymmetry parameter correlates
with type $T$ as $A=0.19+ 0.02T$, put this into the graph, however, the \RR\ varies between 
0.05--0.81 depending on
bin and choice of sub sample and parameter, with the best correlation for unjustified
binning applied. For this reason, we do not quote the errors or mathematical formulation
for how $A$, $C$, or velocity dispersion, vary with type, but take these trends as a
indications and further investigate how scale length depends on these parameters as
morphological type indicators.

In Fig.~\ref{fig:CAsigma}, we plot the scale lengths versus asymmetry, concentration, and
velocity dispersion. Although it is shown that scale length decreases at higher concentration 
parameter, a line-fitting exercise reveals $\mathfrak{R^2} =0.12$, which implies insignificant 
correlation. The velocity dispersion has a higher correlation, $\mathfrak{R^2} =0.37$, but still 
with no strong statistical significance. Moreover, the large scatter of the scale length values 
illustrated in Fig.~\ref{fig:CAsigma} is consistent with the galaxies studied by \citet{deJong96}.

This exercise tells us, firstly, that asymmetry, concentration, and velocity dispersion
only correlate weakly with morphological type, and secondly, that even when using these
parameters as morphological type indicators, there is no strong change in disk scale
length for different galaxy types. Furthermore, we have now been able to use the full 
sample of 30374 galaxies with reliable scale lengths.

\begin{figure*}
\begin{center}
 \includegraphics[width=.99\textwidth]{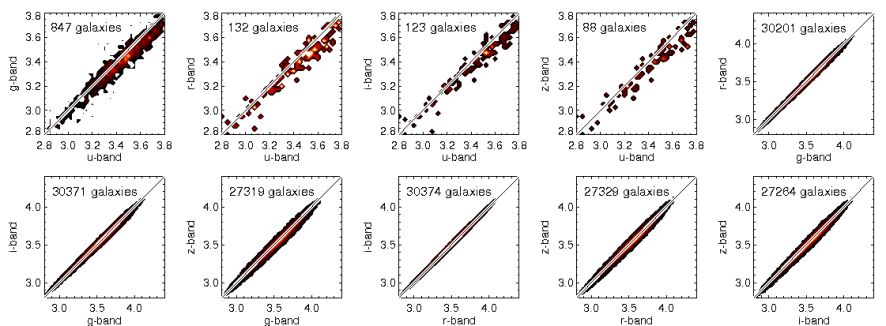}
  \caption{Scale lengths in $u,g,r,i,z$ bands given in decimal logarithm of parsecs, with the number of points in each density plot stated, and the 1:1 line drawn on each panel.} 
\label{fig:rd_ugriz}
\end{center}
\end{figure*}

\subsection{$u, g, r, i,z$ Scale Lengths}
\label{sec:wavelength}
The derived scale lengths can be compared between the images in different bands to
investigate the effects of dust and stellar populations in disk galaxies \citep[e.g., ][]{
Cunow01}. Such analysis is complementary to the results on colour gradients used to analyse
age gradients in disks \citep[e.g., ][ and many more]{ Cunow04}. 

We analyse the derived scale lengths in different bands applying the limits presented in
table~\ref{tab:ugriz}. We can then compare for a given subset, where reliable scale lengths
have been derived in two bands, how the scale length changes between different SDSS bands.
We derive a series of correlations between the scale lengths in different bands, and
although not all band-pair samples are of equal size, we find that the correlations for all
the pairs are significant (see equations (3)--(12), where all formal errors and
coefficients of determination are given).

\begin{eqnarray}
\log(r_d^g) =0.25(\pm0.03) + 0.91(\pm0.01)\log(r_d^u)\ \ \mathfrak{R^2}=0.94\\
\log(r_d^r) =0.36(\pm0.08) + 0.88(\pm0.02)\log(r_d^u)\ \ \mathfrak{R^2}=0.92\\
\log(r_d^i) =0.32(\pm0.07) + 0.89(\pm0.02)\log(r_d^u)\ \ \mathfrak{R^2}=0.94\\
\log(r_d^z) =0.36(\pm0.09) + 0.88(\pm0.03)\log(r_d^u)\ \ \mathfrak{R^2}=0.92\\
\log(r_d^r) =0.06(\pm0.01) + 0.98(\pm0.01)\log(r_d^g)\ \ \mathfrak{R^2}=0.98\\
\log(r_d^i) =0.10(\pm0.01) + 0.97(\pm0.01)\log(r_d^g)\ \ \mathfrak{R^2}=0.98\\
\log(r_d^z) =0.09(\pm0.01) + 0.97(\pm0.01)\log(r_d^g)\ \ \mathfrak{R^2}=0.96\\
\log(r_d^i) =0.04(\pm0.01) + 0.99(\pm0.01)\log(r_d^r)\ \ \mathfrak{R^2}=1.00\\
\log(r_d^z) =0.03(\pm0.01) + 0.99(\pm0.01)\log(r_d^r)\ \ \mathfrak{R^2}=0.98\\
\log(r_d^z) =0.00(\pm0.01) + 1.00(\pm0.01)\log(r_d^i)\ \ \mathfrak{R^2}=0.98
\end{eqnarray}

Although scale lengths derived from different SDSS bands do not show significant
differences, their general trends are as predicted by \citet{Cunow98}. Typically, the
correlations are very strong, and in almost all bands, the corresponding average scale
lengths are comparable:
$\langle r_d^u \rangle = 5.12 (\pm 3.36)$ kpc,\ 
$\langle r_d^g \rangle = 3.85 (\pm 2.10)$ kpc,\ 
$\langle r_d^r \rangle = 3.79 (\pm 2.05)$ kpc,\ 
$\langle r_d^i \rangle = 3.81 (\pm 2.05)$ kpc,\ 
$\langle r_d^z \rangle = 3.75 (\pm 2.02)$ kpc. 

We further find that these numbers are consistent with, e.g., \citet{Courteau96}, 
\citet{deJong96}, and \citet{deGrijs98} who presented an extensive analysis of deep images
of 349, 86, and 45 spiral galaxies, respectively. It should be noted that the sample 
of \citet{deGrijs98} is a sample of edge-on galaxies, which explains the
relatively insignificant variations found in our analysis, as opposed to theirs. Also,
their wavelength range, from B to K, is larger than with SDSS data.   
In an attempt to extend these results to compare with near-infrared
results, we also cross-correlated our sample with the Two Micron All Sky Survey (2MASS) $J,
H, K$-band images. We applied the code presented in section~\ref{sec:scalelengths} and
found the 2MASS images are  too shallow to yield anything presentable.

\begin{figure}
\begin{center}
 \includegraphics[width=.49\textwidth]{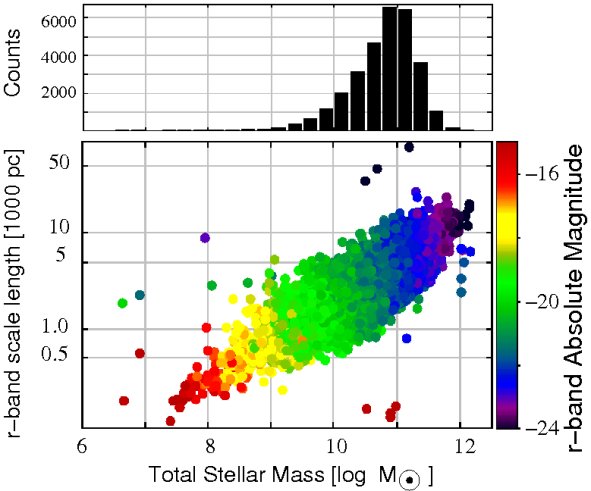}
  \caption{Total stellar mass distribution of our sample is illustrated in the top panel, and the Scale length versus mass diagram shows the expected behaviour in the bottom panel. The colour code shows the $r$-band absolute magnitude derived from the SDSS apparent $r$-band magnitude. The histogram and the scatter plot illustrate the same mass range.} 
\label{fig:rd_mass}
\end{center}
\end{figure}
\subsection{Scale Length versus Stellar Mass}
\label{sec:mass}
We retrieve stellar masses for our sample galaxies by cross matching our sample with the publically available sample of \citet{Kauffmannetal03} and \citet{Brinchmannetal04}, and find stellar masses for 30126 (i.e., almost all our sample) galaxies. These authors have calculated total stellar masses for nearly a million SDSS galaxies based on photometry of the outer regions of the galaxies with models that produce dust corrected star formation rates and use upgraded stellar population synthesis spectra for the continuum subtraction. As noted by these authors, comparison between the stellar masses derived from photometry and spectra from the central regions show that the fits to the photometry are more constrained at low mass since the emission line contribution makes the line index fits less well-constrained. Since the spectroscopic masses are based on fibre spectra from SDSS, which cover only a fraction of the galaxies, we opt to use the stellar masses from photometry. 

In Fig.~\ref{fig:rd_mass} we illustrate the scale length as a function of total stellar mass, and find that larger mass galaxies have larger scale lengths. Moreover, we find that this increase is accompanied by a larger spread in scale length. Galaxies with total stellar mass less than $10^8 \rm{~M}_\odot$ have average $r$-band scale length of $238 \pm 94$ pc,  galaxies with total stellar mass between $10^9$ and $10^{10}\rm{~M}_\odot$ have average scale length of $1.52 \pm 0.65$ kpc, and galaxies with total stellar mass between $10^{11}$ and $10^{12}\rm{~M}_\odot$ have average scale length of $5.73 \pm 1.94$ kpc. All points in Fig.~\ref{fig:rd_mass} are also color coded to show the consistently increasing intrinsic absolute magnitude for larger stellar mass galaxies.

\section{Discussion}
\label{sec:discussion}
We have derived reliable scale lengths for 30374 disk/spiral galaxies, 
with no sign of ongoing interaction or disturbed morphology, in all
five $u,g,r,i,$ and $z$-bands from SDSS DR6 images. Cross-correlation of the SDSS sample
with the LEDA catalogue has enabled us to investigate the variation of the scale lengths
for different types of disk/spiral galaxies. Although the typical scale length in 
$u$-band is 35\% larger than that in the $r$-band, the scale lengths in the 
$g,r,i,$ and $z$-bands are similar and only become smaller on the average for late
morphological types. This result remains consistent
when using by-eye morphological classification or when using asymmetry parameter,
concentration parameter, or velocity dispersion as an indicator for galaxy morphological
type. Our sample spans a range of total stellar masses between $10^{6.6}$ and $10^{12.2} \rm{~M}_\odot$ with a typical galaxy mass of $10^{10.8\pm0.54} \rm{~M}_\odot$, and shows that the while scale length increases for more massive galaxies, the scale length spread also increases with galaxy mass. Overall, these results are in full agreement with the recent work by \citet{Courteauetal07}. 

Scale length variations between bands are commonly studied to better understand the content
and distribution of different stellar populations, metals, and/or dust. A colour gradient
is expected to increase from early-type spirals to late-type spirals,  mainly due
to extinction, which increases to later types. However, for Scd galaxies or later the 
colour gradient becomes smaller, because of decreasing amount of extinction \citep[see, e.g., ][]{PB96,deGrijs98}.

Changes in galaxy scale length in different wavelengths can be attributed to extinction by
moderate amounts of dust, with radial metallicity and age gradients 
as other contributing factors 
\citep[][]{EE84,Petal94, Betal96}. All these parameters will probably change as a
function of redshift, enabling us to measure the variation of intrinsic scale length with 
cosmological epoch. It is expected that the opacity of disk galaxies is expected to have been systematically higher in the
past \citep[e.g., ][]{Dwek98, Pei99}. The fact that the observed radial colour vary
little suggests strongly that stellar population effects are not important here. 
Dust effects are studied by using radiative transfer
models which take into account scattering as well as absorption by dust, and
observationally by investigating the scale length ratio in different bands as a function of
inclination, i.e., optical depth. However, there are some degeneracies. 
Any tendency of the stars in the outer parts of disks to
be bluer would tend to result in underestimated dust content. Any tendency of the dust to concentrate towards the centre would result in an overestimate of the bluer scale lengths, and would not be distinguishable photometrically from a tendency of the stars in the outer disc to be bluer \citep[c.f.][]{PT06,Erwinetal08,Azzolinietal08}. 
\citet{Petal94,Petal95} found
that scale length ratios could change, due to stellar population changes, by a 
factor of approximately 1.1-1.2 from blue to near-infrared ($K$-band). This would correspond to a factor
of about 1.03-1.06 from $g$ to $z$. Since these numbers are very small, a very accurate
analysis is needed to derive conclusions from the SDSS database. Our results seem to first
order in agreement with these numbers.

The derived scale lenghts and our presentation of the transformation coefficients for converting observed scale lengths from one SDSS band to another, furthermore, are meant to be useful tools to test the results of cosmological galaxy formation models, whether numerical, or semi-analytical.

In the future, we plan to add the stored parameters form the Galaxy
ZOO\footnote{http://www.galaxyzoo.org} project to obtain a larger sample with morphological
classifications (at this stage, detailed morphologies are not available), but also compare
scale length as a function environment, nuclear activity, and colour gradients \citep[e.g., comparing the sample with that of ][]{Chatzimi05}. Many more parameters can be further
investigated, and with the data and the derived parameters at hand, we now have the
capability to continue this project in various and potentially unforeseeable directions.

\section*{Acknowledgments}
This work made use of EURO-VO software, tools and services. The EURO-VO has been funded by
the European Commission through contract numbers RI031675 (DCA) and 011892 (VO-TECH) under
the 6th Framework Programme and contract number 212104 (AIDA) under the 7th Framework
Programme. We also acknowledge the use of NASA's SkyView facility
(http://skyview.gsfc.nasa.gov) located at NASA Goddard Space Flight Center, the usage of
the HyperLeda database (http://leda.univ-lyon1.fr), and the TOCAT software
(http://www.starlink.ac.uk/topcat/). KF acknowledges support from the Swedish Research
Council (Vetenskapsr\aa det), and the hospitality of ESO-Garching where parts of this work
were done. KF also acknowledges support from Sergio Gelato for computer support, and fruitful 
discussions with Robert Cumming and Genoveva Micheva. Finally we thank the referee 
Frederic Bournaud for insightful and encouraging comments which helped improve our manuscript.

Funding for the SDSS and SDSS-II has been provided by the Alfred P. Sloan Foundation, the
Participating Institutions, the National Science Foundation, the US Department of Energy,
the National Aeronautics and Space Administration, the Japanese Monbukagakusho, the Max
Planck Society, and the Higher Education Funding Council for England. The SDSS Web site is
http://www.sdss.org/. 

The SDSS is managed by the Astrophysical Research Consortium for the Participating
Institutions. The Participating Institutions are the American Museumof Natural History,
Astrophysical Institute Potsdam, University of Basel, University of Cambridge, CaseWestern
Reserve University, University of Chicago, Drexel University, Fermilab, the Institute for
Advanced Study, the Japan Participation Group, Johns Hopkins University, the Joint
Institute for Nuclear Astrophysics, the Kavli Institute for Particle Astrophysics and
Cosmology, the Korean Scientist Group, the Chinese Academy of Sciences (LAMOST), Los Alamos
National Laboratory, the Max Planck Institute for Astronomy (MPIA), the Max Planck
Institute for Astrophysics (MPA), NewMexico State University, Ohio State University,
University of Pittsburgh, University of Portsmouth, Princeton University, the United States
Naval Observatory, and the University of Washington.

\label{lastpage}

\begin{thebibliography}{} 
\bibitem[Abraham et al.(1996)]{Abrahametal94} Abraham, R.~G., van den 
Bergh, S., Glazebrook, K., Ellis, R.~S., Santiago, B.~X., Surma, P., Griffiths, R.~E. 1996, ApJS, 107, 1 
\bibitem[Adelman-McCarthy et al. (2008)]{AMcetal08} Adelman-McCarthy, J. K. et al. 2008, ApJS, 175, 297
\bibitem[Aguerri et al. (1998)]{Aetal98} Aguerri, J. A. L., Beckman, J. E., Prieto, M. 1998, AJ, 116, 2136
\bibitem[Azzolini et al. (2008)]{Azzolinietal08} Azzolini, R., Trujillo, I., Beckman, J. E. 2008, ApJ, 679, L69
\bibitem[Balcells \& Peletier (1994)]{BP94} Balcells, M., Peletier, R. F. 1994, AJ, 107, 135
\bibitem[Baggett et al. (1998)]{BBA98} Baggett, W. E., Baggett, S. M., Anderson, K. S. J. 1998, AJ, 116, 1626
\bibitem[Beckman et al. (1996)]{Betal96} Beckman, J.~E., Peletier, R.~F., Knapen, J.~H., Corradi, R.~L.~M., Gentet, L.~J. 1996, ApJ, 467, 175
\bibitem[Bertin \& Arnouts (1996)]{BA96} Bertin, E., Arnouts, S. 1996, A\&AS, 117, 393
\bibitem[Brinchmann et al. (2004)]{Brinchmannetal04} Brinchmann, J. et al. 2004, MNRAS, 351, 1151
\bibitem[Boroson (1981)]{Boroson81} Boroson, T. 1981, ApJS, 46, 177
\bibitem[Bournaud et al. (2007)]{Bournaud07} Bournaud, F., Elmegreen, B. G., Elmegreen, D. M. 2007, ApJ, 670, 237
\bibitem[Burda \& Feitzinger (1992)]{BF92} Burda, P., Feitzinger, J. V. 1992, A\&A, 261, 697
\bibitem[Byun \& Freeman (1995)]{BF95} Byun, Y. I., Freeman, K. C. 1995, ApJ, 448, 563
\bibitem[Ceverino et al. (2010)]{Ceverinoetal10} Ceverino, D., Dekel, A., Bournaud, F. 2010, MNRAS, in press, arXiv:0907.3271
\bibitem[Combes \& Elmegreen (1993)]{CE93} Combes, F., \& Elmegreen, B. G. 1993, A\&A, 271, 391
\bibitem[Conselice (1997)]{Conselice97} Conselice, C. J. 1997, PASP, 109, 1251
\bibitem[Conselice (2003)]{Conselice03} Conselice, C. J. 2003, ApJSS, 147, 1
\bibitem[Conselice et al. (2003)]{Conseliceetal03}  Conselice, C.~J., Bershady, M.~A., Dickinson, M., Papovich, C. 2003, ApJ, 126, 1183
\bibitem[Coulais et al. (2009)]{GDL} Coulais, A., Schellens, M., Gales, J., Arabas, S., Boquier, M., Chanial, P., Messmer, P., 2009: Status of GDL - GNU Data Language, Proc. of the 19th conference on Astronomical Data Analysis Software and Systems, Sapporo, Japan
\bibitem[Courteau (1996)]{Courteau96} Courteau, S. 1996 ApJS, 103, 363
\bibitem[Courteau et al. (2007)]{Courteauetal07} Courteau, S. et al. 2007, ApJ, 671, 203
\bibitem[Cunow (1998)]{Cunow98} Cunow, B. 1998, A\&ASS, 129, 593
\bibitem[Cunow (2001)]{Cunow01} Cunow, B. 2001, MNRAS, 323, 130
\bibitem[Cunow (2004)]{Cunow04} Cunow, B. 2004, MNRAS, 353, 477
\bibitem[Dalcanton et al. (1997)]{DSS97} Dalcanton, J. J., Spergel, D. N., Summers, F. J. 1997, ApJ, 482, 659
\bibitem[de Grijs (1998)]{deGrijs98} de Grijs, R. 1998, MNRAS, 299, 595
\bibitem[de Jong (1996)]{deJong96} de Jong, R. S. 1996, A\&A, 313, 45
\bibitem[de Vaucouleurs (1948)]{deV48} de Vaucouleurs, G. 1948, AnAp, 11, 247
\bibitem[de Vaucouleurs et al. (1991)]{deV91}  de Vaucouleurs, G., de Vaucouleurs, A., Corwin, H.~G., Jr., Buta, R.~J., Paturel, G., Fouque, P. 1991, Third Reference Catalogue of Bright Galaxies Volume 1-3, Springer-Verlag: Berlin, Heidelberg, New York
\bibitem[Doi et al. (1993)]{Doietal93} Doi, M., Fukugita, M., Okamura, S. 1993, MNRAS, 264, 832
\bibitem[Dutton (2009)]{Dutton09} Dutton, A. 2009, MNRAS, 396, 141
\bibitem[Dwek (1998)]{Dwek98} Dwek, E. 1998, ApJ, 501, 643
\bibitem[Elmegreen \& Elmegreen (1984)]{EE84} Elmegreen, D. M. \& Elmegreen, B. G. 1984, ApJS, 54, 127
\bibitem[Elmegreen et al. (1996)]{Eetal96} Elmegreen, B. G. Elmegreen, D. M., Chromey, F. R., Hasselbacher, D. A., Bissell, B. A. 1996, AJ, 111, 2233  
\bibitem[Elmegreen et al. (2005)]{Eetal05} Elmegreen, B. G., Elmegreen, D. M., Vollbach, D. R., Foster, E. R., Ferguson, T. E. 2005, 634, 101  
\bibitem[Erwin et al. (2008)]{Erwinetal08} Erwin, P., Pohlen, M., Beckman, J. E. 2008, AJ, 135, 20
\bibitem[Fathi \& Peletier (2003)]{Fathi03} Fathi, K., Peletier, R. F. 2003, A\&A, 407, 61
\bibitem[Fathi (2004)]{Fathi04} Fathi, K. 2004, PhD Thesis, Groningen University
\bibitem[Freeman (1970)]{Freeman70} Freeman, K. C. 1970, 160, 811
\bibitem[Fukugita et al. (2007)]{Fuku07} Fukugita, M. et al. 2007, AJ, 134, 579
\bibitem[Giovanelly \& Haynes (2002)]{GH02} Giovanelli, R., Haynes, M. 2002, ApJ, 571, L107
\bibitem[Governato et al. (2010)]{Governatoetal10} Governato, F. et al. 2010, Nature, 463, 203
\bibitem[Graham (2001)]{Graham01} Graham, A. W. 2001, AMNRAS, 326, 543
\bibitem[Graham \& de Blok (2001)]{GdeB01} Graham, A. W., de Blok, W. J. G. 2001, ApJ, 556, 177
\bibitem[Graham \& Worley (2008)]{GW08} Graham, A. W., Worley, C.C. 2008, MNRAS, 388, 1708
\bibitem[Hatziminaoglou et al. (2005)]{Chatzimi05} Hatziminaoglou, E. et al. 2005, MNRAS, 364, 47
\bibitem[Holwerda (2005)]{Holwerda05} Holwerda, B. 2005, PhD. Thesis, Groningen University
\bibitem[Simien \& de Vaucouleurs (1983)]{SdeV83} Simien, F. \& de Vaucouleurs, G. 1983, IAUS, 100, 375
\bibitem[Lauberts (1982)]{L82} Lauberts A. 1982, The ESO/Uppsala Survey of the ESO(B) Atlas, European Southern Observatory
\bibitem[Lin \& Pringle (1987)]{LP87} Lin, D. N. C., Pringle, J. E. 1987, MNRAS, 225, 607 
\bibitem[Loveday (1996)]{L96} Loveday J. 1996, MNRAS, 278, 1025
\bibitem[Kauffmann et al. (2003)]{Kauffmannetal03} Kauffmann, G. et al. 2003, MNRAS, 341, 33
\bibitem[Kent (1985)]{Kent85} Kent, S. M. 1985, ApJS, 59, 115
\bibitem[Knapen (2004)]{Knapen04} Knapen, J. H. 2004, "Penetrating Bars Through Masks of Cosmic Dust", in ASSL, 319, 189
\bibitem[Knapen \& van der Kruit (1991)]{KvK91} Knapen, J. H., van der Kruit, P. C. 1991, A\&A, 248, 57
\bibitem[Knezek (1993)]{Knezek93} Knezek, P. 1993, PhD thesis, University of Massachusetts
\bibitem[Kormendy \& Kennicutt (2004)]{KK04} Kormendy, J. \& Kennicutt, Jr., R. C. 2004, ARA\&A, 42, 603 
\bibitem[MacArthur (2003)]{MacA03} MacArthur, L. A., Courteau, S., Holtzman, J. A. 2003, ApJ, 582, 689
\bibitem[Martig \& Bournaud (2010)]{MartigBournaud10} Martig, M., Bournaud, F. 2010, ApJ, submitted (arXiv:0911.0891)
\bibitem[Mo et al. (1998)]{Mo98} Mo, H. J., Mao, S., White, S. D. 1998, MNRAS, 295, 319
\bibitem[Mosteller \& Tukey (1977)]{MT77} Mosteller, F., Tukey, J. 1977, Data Analysis and Regression, Addison-Wesley
\bibitem[Nilson (1973)]{N73} Nilson P. 1973, Uppsala General Catalogue of Galaxies, Uppsala Astr. Obs. Annaler, Band 6
\bibitem[Paturel et al. (2003)]{Petal03} Paturel, G., Petit, C., Prugniel, P., Theureau, G., Rousseau, J., Brouty, M., Dubois, P., Cambr{\'e}sy, L. 2003, A\&A, 412, 45
\bibitem[Pei et al. (1999)]{Pei99} Pei, Y. C., Fall, S. M., Hauser, M. G. 1999, ApJ, 522, 604
\bibitem[Peletier et al. (1994)]{Petal94} Peletier, R. F., Valentijn, E. A., Moorwood, A. F. M., Freudling, W. 1994, A\&AS, 108, 621
\bibitem[Peletier et al. (1995)]{Petal95} Peletier, R. F., Valentijn, E. A., Moorwood, A. F. M., Freudling, W., Knapen, J. H., Beckman, J. E. 1995, A\&A, 300, L1
\bibitem[Peletier \& Balcells (1996)]{PB96} Peletier, R. F., Balcells, M. 1996, in "Spiral Galaxies in the Near-IR", ESO/MPA proceedings, eds: Dante Minniti and Hans-Walter Rix, Springer-Verlag Berlin Heidelberg New York
\bibitem[Prieto et al. (2001)]{Petal01} Prieto, M., Aguerri, J. A. L.. Varela, A. M.. Mu\~{n}oz-Tu\~{n}n, C. 2001, A\&A, 367, 405
\bibitem[Pohlen \& Trujillo (2006)]{PT06} Pohlen, M., Trujillo, I. A\&A, 454, 759
\bibitem[Romanishin et al. (1983)]{Romanishin83} Romanishin, W., Strom, K. M., Strom, S. E. 1983, ApJS, 53, 105
\bibitem[Sandage (1961)]{Sandage61} Sandage, A. 1961, The Hubble Atlas of Galaxies, Washington: Carnegie Institute Washington
\bibitem[Schade et al. (1995)]{Schadeetal95} Schade, D., Lilly, S. J., Crampton, D., Hammer, F., Le Fevre, O., Tresse, L. 1995, ApJ, 451, L1
\bibitem[Schaye et al. (2010)]{Schayeetal10} Schaye, J. et al. 2010, MNRAS, 402, 1536
\bibitem[Schombert et al. (1992)]{Schombertetal92} Schombert, J. M. Bothun, G. D., Schneider, S. E., McGaugh, S. S. 1992, AP, 103, 1107
\bibitem[S\'ersic (1968)]{Sersic68} S\'ersic, J. L. 1968, Atlas de galaxias australes, Observatorio Astronomico Cordoba
\bibitem[Shimasaku et al. (2001)]{Shimasaku01} Shimasaku, K. et al. 2001, AJ, 122, 1238
\bibitem[Silk (2001)]{Silk01} Silk, J. 2001, MNRAS, 324, 313 
\bibitem[Valentijn (1990)]{Val90} Valentijn, E. A. 1990, Nature, 346, 153
\bibitem[van der Kruit (1987)]{vdKruit87} van der Kruit, P. C. 1987, A\&A, 173, 59
\bibitem[van Driel et al. (1995)]{vDetal95} van Driel, W. Valentijn, E. A., Wesselius, P. R., Kussendrager, D. 1995, A\&A, 298, 41 
\bibitem[Vorontsov-Velyaminov et al. (1963)]{VV63} Vorontsov-Velyaminov B.A., Arkipova V.P., Kranogorskaja A.A.n 1963-1974, Morphological Catalogue of Galaxies, Trudy Sternberg Stat. Astr.Inst. 32,33,34
\bibitem[Yamauchi et al. (2005)]{Yamauchietal05} Yamauchi, C. et al. 2005, ApJ, 130, 1545
\bibitem[York et al. (2000)]{Yorketal00} York, D. G. et al. 2000, AJ, 120, 1579
\end{thebibliography}
\end{document}